\documentclass[letter]{aa}
\usepackage[dvips]{graphicx}
\usepackage{txfonts}

\begin{document}


\title{Heavy water around the L1448-mm protostar}

\author{C. Codella\inst{1}, C. Ceccarelli\inst{2}, B. Nisini\inst{3}, R. Bachiller\inst{4}, 
J. Cernicharo\inst{5}, F. Gueth\inst{6}, 
A. Fuente\inst{4}, B. Lefloch\inst{2}}

\offprints{C. Codella, codella@arcetri.astro.it}


\institute{
INAF, Osservatorio Astrofisico di Arcetri, Largo E. Fermi 5, I-50125 Firenze, Italy
\and
Laboratoire d'Astrophysique de l'Observatoire de Grenoble, BP 53, 38041 Grenoble Cedex, France
\and
INAF - Osservatorio Astronomico di Roma, via di Frascati 33, 00040 Monteporzio Catone, Italy
\and
Observatorio Astr\'onomico Nacional (IGN), Alfonso XII, E-28014 Madrid, Spain
\and
Laboratory of Molecular Astrophysics, CAB (CSIC-INTA), Crta Ajalvir km 4, E-28840 Madrid, Spain
\and
IRAM, 300 Rue de la Piscine, F-38406 Saint Martin d'H\`eres, France}

\date{Received date; accepted date}

\titlerunning{HDO towards the protostar L1448-mm}
\authorrunning{Codella et al.}

\abstract{L1448-mm is the prototype of a low-mass Class 0 protostar
  driving a high-velocity jet. Given its bright H$_2$O spectra
  observed with ISO, L1448-mm is an ideal laboratory to observe
  heavy water (HDO) emission.}  
{Our aim is to image the HDO emission in the protostar surroundings,
  the possible occurrence of HDO emission also investigating off
  L1448-mm, towards the molecular outflow.}  
{We carried out observations of L1448-mm in the HDO(1$_{\rm
    10}$--1$_{\rm 11}$) line at 80.6 GHz, an excellent tracer of HDO
  column density, with the IRAM Plateau de Bure Interferometer.}  
{We image for the first time HDO emission around L1448-mm.  The HDO
  structure reveals a main clump at velocities close to the ambient
  one towards the the continuum peak that is caused by the dust heated by the
  protostar.  In addition, the HDO map shows tentative weaker emission 
  at $\simeq$ 2000 AU from the protostar towards the
  south, which is possibly associated with the walls of the
  outflow cavity opened by the protostellar wind.}  
{Using an LVG code, modelling the density and
temperature profile of the hot-corino, and adopting a gas temperature of
100 K and a density of $1.5 \times 10^8$ cm$^{-3}$, we derive a beam diluted HDO column
density of $\sim$ 7 $\times$ 10$^{13}$ cm$^{-2}$,
corresponding to a HDO abundance of $\sim$ $4 \times 10^{-7}$.
In addition, the present map supports the scenario where HDO can be efficiently
produced in shocked regions and not uniquely in hot corinos heated
by the newly born star.} 


\maketitle

\section{Introduction}

Water can be considered as the most important
molecule in the oxygen chemistry (after CO), influencing in particular
the overall chemical composition of the gas involved in the process of
star formation. In addition H$_2$O is the most abundant component in
the grain mantles and plays a fundamental role in the energy balance
of star-forming regions (e.g., Ceccarelli et al. \cite{cecca96}; Kaufman
\& Neufeld \cite{kaufman}; Bergin et al. \cite{bergin98}; van Dishoeck
\& Blake \cite{vanblake}).  In starless regions water abundance is
very low ($\le$ 10$^{-8}$, e.g., Bergin \& Snell \cite{berginsnell}),
while it increases to the abundance of CO ($\le$ 10$^{-4}$) near
protostars (e.g. Liseau et al. \cite{liseau}; Harwit et
al. \cite{harwit}).  Several mechanisms have been proposed for H$_2$O
formation (see the recent work on H$_2$$^{18}$O 
by J{\o}rgensen \& van Dishoeck \cite{jorgevan}, and
reference therein): (i) heating from forming stars, (ii) jet-driven
shocks, or (iii) shocks in accretion disks. In particular, water
emission becomes even more important when studied in low-mass
protostars, which are possibly associated with solar-type
protoplanetary systems.

Unfortunately, the H$_2$O low-energy lines are almost invisible from
the ground due to very strong atmospheric absorption, and only under
very special conditions are they observed towards the brightest sources
(Cernicharo et al. \cite{pepe90}). 
As an example, H$_2$O emission has been
detected towards the low-mass protostar NGC1333-IRAS4B
by different facilities, namely ISO, Odin, SWAS, Spitzer and, very recently,
Herschel (Maret et al. \cite{maret}; Bjerkeli et al. \cite{bjerkeli}; Bergin et al. \cite{bergin03};
Franklin et al. \cite{franklin}; Watson et al. \cite{watson}; Kristensen et al. \cite{kristensen}).
These observations, with their relatively large apertures, revealed water from
different components, namely the protostar, the outflow and the extended ambient medium.
Recently  J{\o}rgensen \& van Dishoeck (\cite{jorgevan}) imaged IRAS4B in the mm-range with the PdB
interferometer. They detected H$_2$O emission associated with the protostar,
probably coming from the inner 25 AU of
the accretion disk, which calls for further observations of low-mass
objects.

Given the difficulty of detecting water, several alternatives have
been pursued. The most natural one is to observe rare isotopologues
like HDO, which suffer much less telluric absorption and have some
lines at sub-mm and mm-frequencies. Furthemore, if HDO is formed on the
surfaces of grains before the heating due to the radiation from an
accreting protostar (e.g., Parise et al. \cite{parisea}, and
references therein), then its emission can also be used to trace the
past history of water formation. Heavy water observations have been
successfully carried out towards high-mass star-forming regions
(Plambeck \& Wright \cite{plambeck}; Jacq et al. \cite{jacq};
Gensheimer et al. \cite{gensheimer}; Helmich et al. \cite{helmich};
Pardo et al. \cite{pardo}; Comito et al. \cite{comito}), but they are
still very sparse.  In Orion, the emission 
arises from warm gas ($\rm T_k$ $>$ 150 K) and the abundance of HDO is
strongly enhanced ($>$ $10^3$) with respect to ambient values, 
most likely because of recent HDO evaporation
from dust grain mantles (see also Bergin et al. 2010, in the 
HIFI A\&A special issue). 
On the othe hand, the number of studies of HDO towards the
lower-mass star-forming regions is even less.  Stark et
al. (\cite{stark}) and Parise et al. (\cite{parisea}) observed HDO
emission with the JCMT and IRAM single-dish antennas, towards the
hot corino heated by the IRAS16293 protostar. In this case the
measured deuteration ratio (3 10$^{-2}$) supports models where water
is produced in the gas phase at low temperature and then storaged on
the grains (e.g., Roberts et al. \cite{roberts}). On the other hand,
if H$_2$O is formed in shocks, followed by condensation out on the
grain mantles, the deuteration ratio is predicted to be lower than
10$^{-3}$. It is still missing a study of HDO emission through
interferometric high angular resolution images, which
is needed to distinguish the hot-corino emission from that
associated with more extended outflows.

In this Letter, we report for the first time a high-resolution image
of HDO at 80.6 GHz towards the protostar L1448-mm. We show that the
emission at velocities close to the systemic one comes from the hot
corino, we also report tentative HDO emission offset from the
protostar, which is possibly associated with the cavity of the outflow red
lobe.

\section{Observations and source selection}

L1448-mm, located at $\sim$ 250 pc from the Sun, can be considered as
the prototype of a low-luminosity ($\sim$ 8 $L_{\rm \odot}$) Class 0
protostar.  The newborn star is driving a highly collimated molecular
jet consisting of well separated extremely high-velocity ($\ge$ 50 km
s$^{-1}$) molecular bullets, well detected in SiO and CO (Guilloteau
et al. \cite{guillo92a}; Bachiller et al. \cite{bachiller95}). Lower
velocity CO emission delineates a biconical cavity, possibly created
by the passage of the jet, which is entraining the surrounding gas
through successive bow-shock well detected in H$_2$ (e.g., 
Davis \& Smith \cite{davis}; Dionatos et
al. \cite{dionatos}; Neufeld et al. \cite{neufeld09}).  L1448-mm has
been selected because of its bright H$_2$O spectra that are observed with the ISO
satellite, which lead to a large water abundance, 5 10$^{-4}$, towards
the protostar (Nisini et al. \cite{nisini00}).  The large ISO beam
(80$\arcsec$) does not allow one to exactly localise the H$_2$O-rich
gas, but evidence is given that the high water abundance is maintained on a high
level along the southern red-shifted part of the outflow.

The HDO observations of L1448-mm were obtained in July 2009 with the
IRAM Plateau de Bure Interferometer (PdBI) in France.  Four tracks in
the five-element D configuration (baselines from 24~m up to 94~m) were
used, for a total time of $\sim$19 hours. A region of about 60$\arcsec$
around the protostar has been mapped. 
The HDO(1$_{\rm 10}$--1$_{\rm 11}$) line at 80578.295 MHz was observed with both 20
and 40 MHz bandwidths and a corresponding spectral resolution of
$\sim$ 0.13 and 0.25 km s$^{-1}$, respectively.  Two other units
with 320 MHz bandwidth have been used to measure the continuum. We used the same
setup for vertical and horizontal polarisation to gain in sensitivity.
Amplitude and phase were calibrated by observing 0130--171, 0234+285
and 0333+321, whereas the flux density scale was derived by observing
3C\,84 and 3C\,454.3, with an uncertainty of $\sim$ 20\%.  Images were
produced using natural weighting, and restored with a clean beam of
$6\farcs29\times4\farcs87$ (PA=123$^\circ$).
Note that we observed HDO(1$_{\rm 10}$--1$_{\rm 11}$) emission with a deep
integration (230 minutes on-source) using the 30-m IRAM single-dish antenna
and a HPBW of 31$\arcsec$ (see Sect. 3). 

\section{Results: the HDO maps}

\begin{figure}
\begin{center}
\centerline{\includegraphics[angle=-90,width=8cm]{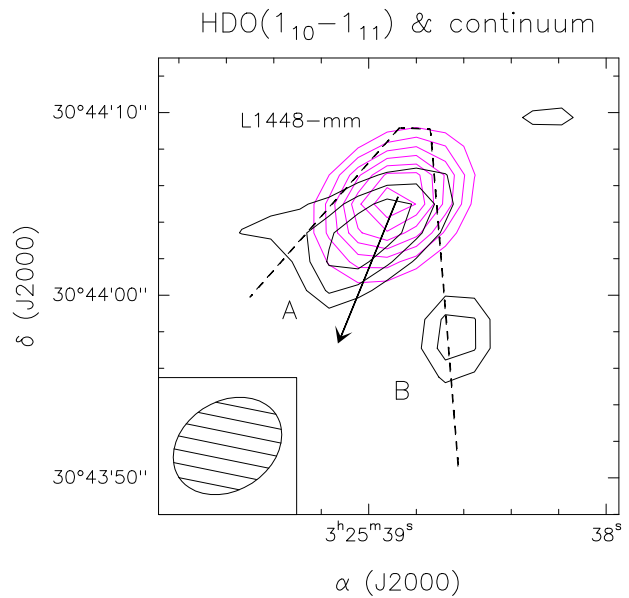}}
\caption{Contour map of the HDO(1$_{\rm 10}$--1$_{\rm 11}$) emission
  (black contours) integrated between --0.3 and +10.7 km
  ss$^{-1}$ superimposed onto the continuum image at 3.7mm
  (magenta). The HDO contour levels range from 3$\sigma$ (33 mJy km s$^{-1}$ 
  beam$^{-1}$) to 5$\sigma$ by 1$\sigma$. Continuum contour levels
  range from 30 ($\sim$ 20$\sigma$) to 90\% of the peak value (16
  mJy), and identify the driving protostar L1448-mm.  The filled
  ellipse in the lower left corner shows the synthesised PdBI beam
  (HPBW): $6\farcs29\times4\farcs87$. The black arrow points in the
  direction of the jet (PA=--21$^\circ$) identified by Guilloteau et
  al. (1992) through SiO emission. The dashed lines identify the sketch
  of the limits of the conical blue-shifted (+6,+8 km s$^{-1}$) cavity
  observed in CO(1--0) at PdBI by Bachiller et al. (1995).}
\label{hdo}
\end{center}
\end{figure}

\begin{figure}
\begin{center}
\centerline{\includegraphics[angle=-90,width=5cm]{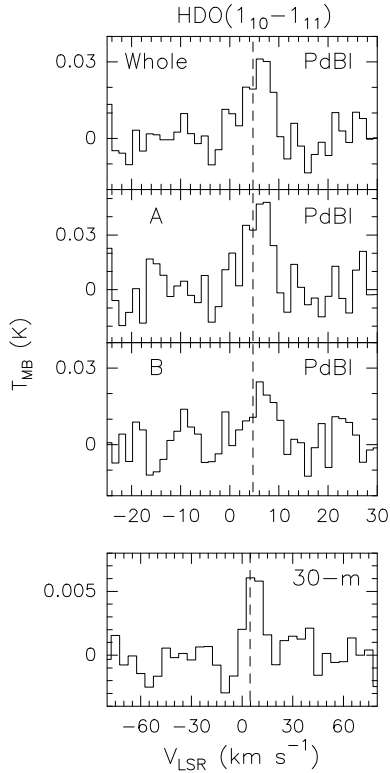}}
\caption{HDO(1$_{\rm 10}$--1$_{\rm 11}$) spectra (in $T_{\rm B}$
  scale) derived from the PdBI image and integrated over  
  the whole emitting region (upper panel), and
  the area of the two structures called A and B
  (middle panels). The spectral resolution is 1.4 km s$^{-1}$. Bottom
  panel reports the HDO(1$_{\rm 10}$--1$_{\rm 11}$) low-spectral
  resolution (5 km s$^{-1}$) spectrum as observed with the 30-m IRAM
  antenna. The
  ambient LSR velocity (+4.7 km s$^{-1}$) is marked by the vertical
  dashed line.}
\label{spectra}
\end{center}
\end{figure}

Figure \ref{hdo} shows the contour map of the HDO(1$_{\rm
  10}$--1$_{\rm 11}$) integrated emission (contours) towards L1448-mm.
The magenta contours trace the unresolved 3.7~mm continuum emission with a
peak flux of 16 mJy centred at $\alpha_{\rm J2000}$ = 03$^{\rm h}$
25$^{\rm m}$ 38$\fs$9, $\delta_{\rm J2000}$ = +30$\degr$ 44$\arcmin$
05$\farcs$1, in agreement with previous observations (e.g. Guilloteau
et al. \cite{guillo92b}; Bachiller et al. \cite{bachiller95}).  The
HDO structure reveals a main clump (hereafter called A),
(i) located close to the continuum peak due to the dust heated by the
protostar and (ii) sligthly elongated 
in the S-E direction, i.e. the direction of the red-shifted
outflow lobe. 
The offset seen in Fig. \ref{hdo} between the the clump A and 
the continuum ($\le$ 2$\arcsec$) is marginally significant, since the position
uncertainty is around 1$\arcsec$.  
Although we cannot exclude a possible association of
the clump A with the outflow (see below), we conservatively assume 
that clump A is tracing the hot-corino around the protostar.
Beside clump A, the HDO map shows tentative emission from a weaker
position (here called clump B) located 
at $\sim$ 8$\arcsec$ ($\simeq$ 2000 AU at 250
pc) from the protostar, again towards the southern direction. 
No hints of HDO emission have been detected towards the N-W direction,
where the blue-shifted molecular lobe is located.

Figure \ref{spectra} shows the HDO(1$_{\rm 10}$--1$_{\rm 11}$)
spectra, obtained from the PdBI maps by integrating over the whole
area where heavy water emission is detected (upper panel) and over
clumps A and B (middle panels). In particular, 
a Gaussian fit of the clump A spectrum leads to a
peak of 49 mK (in $T_{\rm MB}$ scale), observed with a S/N = 5, at
+6.0 $\pm$ 0.4 km s$^{-1}$, i.e. slightly red-shifted with respect to
the systemic velocity.  
The integrated intensity is 286 $\pm$ 50 mK km
s$^{-1}$, while the FWHM linewidth is 5.7 $\pm$ 1.0 km s$^{-1}$. The
spectra, smoothed to 1.4 km s$^{-1}$ resolution to increase
sensitivity, do not allow us a careful analysis of the line profile.
Bottom panel of Fig. \ref{spectra} reports the HDO spectrum as
observed at low-spectral resolution (5 km s$^{-1}$) after 230 minutes
on-source using the 30-m IRAM antenna. The HDO line peaks at $\sim$
6 mK (1 $\sigma$ = 0.2 mK).  The line flux is 305 mJy km
s$^{-1}$, showing that the PdB image collects about 61\% (187 mJy km
s$^{-1}$ integrated over the 30-m beam) 
the of the flux
observed with the IRAM single-dish. Given the calibration
uncertainties of the PdBI image ($\sim$ 20\%) and of the 30-m
measurement ($\le$ 15\%), the most plausible solution is that a weak
extended structure has been filtered out by the interferometer.

We can compare the present HDO image with the CO(1--0) map obtained at
the PdBI with a spatial resolution of $\sim$ 3$\arcsec$ by Bachiller
et al. (\cite{bachiller95}, see their Fig. 5).  The dashed lines in
Fig. \ref{hdo} identify the position of the walls of the conical
outflow cavity as observed at red-shifted velocities from +6 to 
+10 km s$^{-1}$). The black arrow shows the direction of the
protostellar jet identified by Guilloteau et al. (1992) through SiO
emission. Considering also the HDO spectrum of clump B, which shows
emission at the same velocities, it is tempting to associate 
the HDO emission tentatively observed offset from the
L1448-mm protostar for the first time with the cavity opened by the jet and well observed
in CO.  Interestingly, one possible origin for the H$_2$O masers at 22
GHz observed at high-angular resolution studies towards low-mass
protostars is its association with conical shells (Marvel et
al. \cite{marvel}), which may be caused by propagation of large bow-shocks
that entrain ambient medium at long transverse distances from the jet
but which could also represent walls of evacuated cavities
produced by a wide-angle wind (see e.g., Shang et al. \cite{shang},
and references therein). The water maser spots require extremely
high-density ($\sim$ 10$^{7}$--10$^{9}$ cm$^{-3}$, Elitzur et
al. \cite{elitzur}, Kaufman \& Neufeld \cite{kaufman}) conditions,
which could be found in compressed shocked layers at the interface
with the surrounding medium. Indeed, Cernicharo et al. (\cite{pepe96})
reported a spectrum at 183 GHz observed with the 30-m IRAM antenna, which 
was owing to the H$_2$O(3$_{\rm 13}$--2$_{\rm 20}$) transition. The H$_2$O
spectrum is composed by several narrow red-shifted components, 
resembling that classical maser spectral pattern. The brightest
feature is at $\sim$ +9 km s$^{-1}$, i.e. close the velocity observed
with the present HDO emission, but the whole pattern covers a 
broad range of velocities, questioning a possible association of the 
183 GHz emission with the cavity walls.

In conclusion, the present HDO map supports the presence of water in
the cavity walls and suggests that HDO can be found not only in heated
hot cores/corinos, but also in shocked material.

\section{The HDO abundance inside and outside the hot corino}

To derive first the HDO column density in clump A and then the
abundance, we used an LVG code with the collisional coefficients from 
Green (\cite{green}), 
in a plane parallel slab (where because the line is very thin optically the
exact geometry is negligible). We considered a gas temperature
in the 50-120 K range and a density between $1\times10^7$
cm$^{-3}$ and $1\times10^9$ cm$^{-3}$.  Figure 3 shows the 
beam-averaged HDO column density corresponding to the signal observed in
the direction of clump A as a function of the gas temperature for
various densities. Within the 50-120 K interval of gas
temperatures, the HDO total column density ($N_{\rm tot}$) varies
very little as function of the gas temperature, and mostly depends
on the gas density. At 100 K, varying the density from $\sim10^8$
to $\sim10^7$ cm$^{-3}$, corresponds to a factor $\sim$ 4 in the HDO column
density. At higher densities, the LTE solution is approached and the
dependence on the gas density is not as strong.

If one assumes that the unresolved HDO emission towards the source is
caused by the sublimation of the water ice mantles when the dust
temperature reaches 100 K, one can derive the abundance of the
sublimated heavy water. We used the density and temperature profiles
of the envelope of L1448-mm derived by J{\o}rgensen et al.
(\cite{jorge}) in order to evaluate the sizes and density of the
sublimation region and its H$_2$ column density. The sublimation
region is predicted to have a radius of $\sim$ 25 AU, equivalent to a
diameter of about 0$\farcs$2. At that radius, the H$_2$ density is
equal to $1.5 \times 10^8$ cm$^{-3}$, giving an H$_2$ column density of
the sublimation region of about $1.1 \times 10^{23}$ cm$^{-2}$. 
The beam-averaged HDO column density derived from the Figure 3,
assuming a density of $1.5 \times 10^8$ cm$^{-3}$ and a temperature
of 100 K, is $\sim 7 \times 10^{13}$ cm$^{-2}$. Thus, taking into
account the dilution factor of the PdBI beam, the HDO
abundance in the hot corino results in $\sim 4 \times 10^{-7}$. 
The main uncertainty in this value is due to the uncertainty in the
radius, density and H$_2$ column density of the sublimation
region. A density a factor 3 larger would decrease the estimated
HDO abundance by less than a factor 2, whereas a factor 3 lower
density would increase the abundance by about a factor 4.
Taking the worst scenario, i.e. that the emission arises in a region
of 5$\farcs$5 size, and using again the structure by J{\o}rgensen
et al. (2002), the emitting region would have a radius of $\sim 700$
AU, where the gas temperature is $\sim 22$ K and the density is
$\sim 1.5\times 10^6$ cm$^{-3}$. 
The HDO column density ($\sim 1 \times 10^{16}$ cm$^{-3}$) and the
abundance ($\sim 4\times 10^{-7}$) would be outrageously large for
such a cold temperature, where water would be largely frozen onto
the grain mantles, and we accordingly exclude this possibility.

The linewidth may bring some additional information. The observed HDO
linewidth is around 6 km s$^{-1}$. Again, following J{\o}rgensen et
al. (\cite{jorge}), if the emission originates in the assumed 25 AU,
the mass of the central object is about 0.1
M$_{\odot}$. Alternatively, if the central object has a mass of 1
M$_\odot$, the emission must originate in a region at about 100 AU,
where the gas temperature is $\sim 50$ K, the density is $\sim
2 \times 10^7$ cm$^{-3}$, and the H$_2$ column density is $6 \times 
10^{22}$ cm$^{-2}$. The beam-averaged HDO column density would be
$5 \times 10^{14}$ cm$^{-2}$ (Fig. 3) and consequently the HDO abundance  
$\sim 5 \times 10^{-7}$.

\begin{figure}
\begin{center}
\centerline{\includegraphics[angle=0,width=8cm]{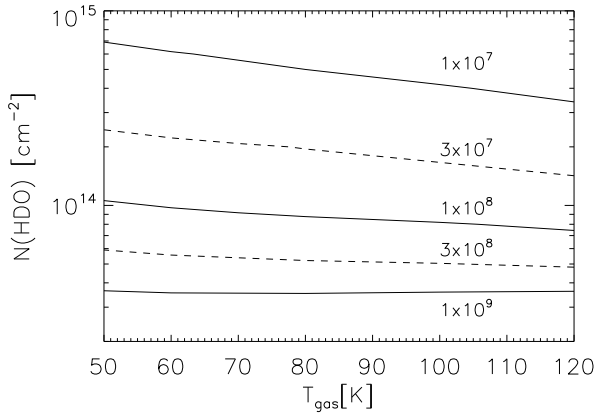}}
\caption{Beam-averaged HDO total column density corresponding to the
(1$_{\rm 10}$--1$_{\rm 11}$) signal observed in the direction of the
hot corino (clump A) as function of a gas temperature T$_{\rm gas}$ between
50 and 120 K and for a
density between $1\times10^7$ cm$^{-3}$ and $1\times10^9$ cm$^{-3}$.
An LVG code  in a plane parallel slab has been used (see Sect. 4).}
\label{model}
\end{center}
\end{figure}

In summary, the present observations suggest that the HDO
abundance is $\sim$ $4 \times 10^{-7}$ and likely originates in a
region whose radius is not much larger than about 100 AU. The most
probable hypothesis is that the HDO emission observed in clump A
is associated with the hot corino, and the sublimation of the icy
mantles.

We also compared our observations with the theoretical predictions by
Parise et al. (\cite{pariseb}), who carried out a theoretical study of
the HDO emission in L1448-mm. Their Fig. B2, upper right panel, shows
the dependence of the HDO flux as function of the HDO abundance in the
outer envelope (where the dust temperature is less than 100 K) and in
the hot corino. The curves are degenerate, in the sense that the
observed signal is obtained either when (1) the HDO abundance is $\sim
2 \times 10^{-8}$ in the outer envelope and lower than $\sim 5 \times
10^{-8}$ in the hot corino or (2) the HDO abundance is $\sim 5 \times
10^{-7}$ in the hot corino and lower than $\sim 3 \times 10^{-9}$ in
the outer envelope. The present interferometric observations suggests
that the emission observed towards L1448-mm originates in the hot
corino, thus solving the degeneracy.  Also, note that the
abundance derived in our simple analysis agrees very well with the
abundance derived using this more sophisticate modelling. 

As discussed in the introduction, HDO can be a useful probe of the
corresponding main isotopologue, H$_2$O, which is largely undetectable from
ground. The study of the water deuteration in a similar solar type
protostar, IRAS16293-2422, by Parise et al. (\cite{parisea}) suggested
that the HDO/H$_2$O ratio in the sublimated ices is around 3\%.  When
comparing L1448-mm and IRAS16293-2422 with other sources where the
molecular deuteration of formaldehyde and methanol was measured, it
appears that both L1448-mm and IRAS16293-2422 are typical sources
(Parise et al. \cite{parise06}). It is then plausible that the
HDO/H$_2$O in L1448-mm is similar to what was found in IRAS16293-2422,
namely $\sim$ 3\%, implying a H$_2$O abundance of about
$\sim 1 \times10^{-5}$. This is somewhat similar to the water abundance
estimated in IRAS16293-2422 by ISO observations (Ceccarelli et
al. \cite{cecca2000}; Crimier et al. \cite{crimier}), and troubling
because it would imply either an abundance of the water ices lower
than usually assumed, $\sim 10^{-4}$, or that not all ices have been
sublimated. New observations by the just launched Herschel satellite
will soon confirm or disregard this result and clarify the
situation.

The HDO emission tentatively observed outside the hot corino and
presumibly associated with the outflow red-shifted cavity observed in
CO is only a factor 2 weaker than that observed towards the protostar
(see HDO spectra of clumps A and B reported in Fig. 2).  Therefore, as
for the hot corino region, this HDO signal corresponds to a HDO column
density of about $10^{14}$ cm$^{-2}$ if the density is around $10^8$ 
cm$^{-3}$ and the emission fills up the PdB beam. It is difficult
to estimate the abundance of HDO because of the lack of available
estimates of the H$_2$ emission in the cavity. Even though we are
unable to quantify how much water has been sputtered or sublimated
from the grain mantles, clearly one or more shocks are responsible for
the HDO spot that is coincident with the cavity wall. In other words, HDO
emission cannot be just limb-brightned emission: water must have been
extracted from the grain mantles by some energetic process. As
previously noted by other authors in other contexts, models of bow
shock naturally would explain the morphology and the presence of shock
``far away'' from the high-velocity jet.

\section{Conclusions}

The present high spatial resolution map of L1448-mm allows us to image
for the first time HDO emission at 80.6 GHz around a Class 0 protostar
with a spatial scale of 5$\farcs$5, i.e. $\sim$ 1400 AU at 250 pc.  We
have shown that the bulk of the emission at velocities close to the
ambient one likely comes from the hot corino, whereas a tentative
detection of HDO emission is present at $\sim$ 2000 AU from the
protostar, and it is possibly associated with the walls of the outflow
cavity previously observed in CO. These findings support the idea that
HDO can be efficiently produced in shocked regions and not uniquely in
hot cores heated by the newly born star.

We derived the HDO abundance in the hot corino with the
reconstruction of the density and temperature profiles by
J{\o}rgensen et al. (\cite{jorge}).  Adopting a gas temperature of
100 K and a density of $1.5 \times 10^8$ cm$^{-3}$, a HDO column
density of $\sim$ 7 $\times$ 10$^{13}$ cm$^{-2}$ was derived,
corresponding to a HDO abundance of $\sim$ $4 \times 10^{-7}$. If we
assume that the HDO/H$_2$O in L1448-mm is similar to what was found in
IRAS16293-2422 ($\sim$ 3\%), a H$_2$O abundance of about
10$^{-5}$ is derived.  
Only future estimates of the H$_2$ column density will allow us to
give an estimate of the HDO abundance in the outflow.

\begin{acknowledgements}
We thank S. Cabrit, M. Tafalla, and J. Santiago-Garc\'{\i}a for 
fruitful discussions and suggestions. We also thank an anonymous referee for 
detailed comments that have greatly helped to improve this paper.
\end{acknowledgements}






\end{document}